\documentclass{article}
\usepackage{url} 
\usepackage{cite}
\usepackage{amsmath,amssymb,amsfonts}
\usepackage{algorithmic}
\usepackage{graphicx}
\usepackage{textcomp}
\usepackage{xcolor}
\def\BibTeX{{\rm B\kern-.05em{\sc i\kern-.025em b}\kern-.08em
    T\kern-.1667em\lower.7ex\hbox{E}\kern-.125emX}}

\begin{document}

\title{Power laws in code repositories: A skeptical approach
  \thanks{This paper has been supported in part by project DeepBio
    (TIN2017-85727-C4-2-P)} }

\author{ Bartolom\'{e} Ortiz\\
  Geneura Team, ETSIIT and CITIC \\
    \textit{University of Granada}\\
    Granada, Spain \\
    bortiz@ugr.es
  \and
  J. J. Merelo\\
  Geneura Team, ETSIIT and CITIC \\
    \textit{University of Granada}\\
    Granada, Spain \\
    jmerelo@ugr.es}

\maketitle

\begin{abstract}
  
  Software development as done using modern methodologies and source
  control management systems, has been often established as an example of
  self-organization, with code growing and evolving organically,
  through activities that do not stem from centralized power, leader
  or directives.  The main challenge in proving these claims is that
  self organization cannot be detected through direct observation, but
  through measurements on the system, looking for hints such as the
  existence of power laws over some features, such as the size of
  changes over time.  The problem we intend to tackle in this paper is to establish a methodology for checking, for a chosen set of repositories we had already measured
  in the past, if the claims about power laws actually hold from a
  precise mathematical point of view, since, although shown as
  pervasive in the software engineering literature (and others), power
  laws are more elusive than they might seem at first sight. For that
  reason, in this paper we present a statistically accurate set of
  tests that will help us decide, from the way repositories are
  changing, if they are really distributed by a power law, which could
  indicate us the existence of a state reached via self-organization,
  or actually, how accurately a power law fits the observed
  distribution of the size of changes of commits in git repositories
  of 16 open source repositories.  We revisit one of the most
  representative papers of these observations to reevaluate its
  results and compare them with the current status of the repositories
  analyzed in it, trying to elucidate if there has been any change in
  the possible presence, or not, of a power law.

\end{abstract}

\textbf{Keywords}: Complex systems, self-organizing systems, self-organized
  criticality, power laws, software development, software repositories

\section{Introduction}\label{introduction}

There are several steps in the study of self-organized systems
\cite{bak1988self}. One of them is to study the existence of a
power law distribution in the data from our system. This property is
intrinsically related to the scale-free behaviour these systems
exhibit, that is, systems where there is no single characteristic
event size \cite{golyk20self}.  Understanding what distribution is
producing our data gives us insights of how confident we are about
claiming that our system is in a critical state, since these states
are usually correlated with the appearance of a power law
\cite{newman2005power}.

Moreover, it is interesting to see if there is some kind of evolution
in the distribution of the data.  A change in the distribution
could be originated by a phase change, which usually implies that
we have crossed a momentary critical state described by a power law.

However, during the process of investigating the existence of self
organization, or the change to a critical state, we have to justify
every conclusion we get, not only by fitting our data, but by giving a
proper statistical measure of the probability of our conclusions, that
is, following an standardize mathematical method based on hypothesis
testing.

In our case, we are mainly interested in code repositories. The
existence or not of a critical state is essential to understand the
processes taking place in free software repositories, which are
sometimes partly or fully based on volunteer work. In these cases,
sustainability is the main objective, and understanding how
self-organization relates to productivity and to the continuous
collaboration of volunteers is essential to keep a software team
healthy. Also, understanding the micro-mechanisms through which
self-organization takes place provides insights on the possible
incentives to avoid churn of volunteers, as well as to make their work
add as much value as possible to the project.

Following the approach outlined in 
\cite{merelo16:self,Merelo2016:repomining,merelo17vue,DBLP:conf/iwann/MereloCV17,merelo2017self}, we are going to try and comprehend code repositories by the
interactions (events) made in them. Code repositories are one of the
main ways software is developed nowadays. Multiple platforms like
GitLab or GitHub offer the possibility of interact with the code
making changes, adding and deleting lines.  All these interactions are
reflected in an archive by the so called commits. From this archive we
can obtain the data about how this code has been changed. Although
these changes sometimes obey issues with the code, or bug fixes, in
many cases they simply are piece-wise improvements or code changes that
are decided by programmers themselves.

But to really start to understand the behavior and possible
self-organization of software development teams as reflected by their
activity in repositories, we should not only try to fit a power law in
our data but try to make a statistically accurate study to confirm how
much evidence we have of this type of behavior.

Most of the measures and estimations about power laws in code
repositories are collected in studies like \cite{merelo2017self},
where some evidences of power laws are presented.  However, since
there are better methods to study this data \cite{clauset2009power},
we are going to re-evaluate all the results obtained, offering an
alternative skeptic view of the existence of power laws, following
recent papers such as \cite{Holme2019,Broido2019}.

We have focused in this particular topic because of three
reasons.\begin{itemize}
\item First of all, there are no extensive analysis that addresses
  this question. We haven't found a comparative study that offers
  data-based statistical measures to check if the power law
  assumptions hold.  So we are going to fill this gap offering a
  complete statistically study about the evidences of power laws
  behavior in a variety of repositories.
\item Moreover, we are not going to evaluate just our evidences about
  the existence of a power law, we are going to offer a short range of
  distributions that can generate similar results and could easily be
  confounded by a power law. This is an standard way to discover if
  our data is more likely to be produced by other heavy-tailed
  distribution \cite{clauset2009power}, like the logNormal or the
  truncated power law.
\item Finally, since we are able to access to multiple years' data, we
  also interested in study the temporal evolution of the two previous
  analysis. Here our main objective is to show how these measure could
  have changed (if they have at all). We should remember that
  repositories are evolving system by definition, so they can
  experiment qualitative changes in his events
  distribution. Furthermore, even if they preserve their behaviour,
  they can still change some distribution's parameters.  To locate and
  analyze this changes could be translated in a deeper knowledge in
  the evolution of these system, specially in the presence of phase
  changes \cite{merelo2017self} or self-organized critically
  (understood as a mechanism by which the system is regulating himself
  to stay in those parameters \cite{newman2005power}).
\end{itemize}

This way, we offer a test to analyze how much evidence there is for or
against our assumptions in every step, and a set of comparatives to
identify some relevant temporal changes, establishing a statistically
correct result about the existence of power laws in repositories.

All these questions lead us to our final result: power laws are not the only concept that should be used to address the question of self organization
in repositories \cite{alderson2010contrasting}, we should use additional measures to assure our conclusions. Even if we use
them, the statistical significance of these fits is very low (which
means that other distributions have at least the same probability of being
generating the data).


Next, we will explain our methodology followed by the results
obtained, closing with our conclusions.


\section{State of the art}\label{soa}

The study of code repositories from the point of view of
self-organized criticality has not followed a continuous line of
investigation. Methods and conclusions often change from one research paper to another,
making difficult to compare results and possible implications. 

On the one hand, we find that self-organized criticality has been
described by a wide variety of code repositories with a certain
probability. These studies can be found
in\cite{wu2007empirical,gorshenev2004punctuated}, but they are not the
only ones of this kind. Tests on the possible self-organized
criticality have also been sought in another type of repositories and
projects, which are enumerable in size and purpose
\cite{Merelo2016:repomining,merelo16:slash,merelo16:self,merelo2017self}.

In most of these studies, however, the evidences in favor of the
existence of a power law are not accompanied by a significant
statistical study. It is usual, as described in \cite{newman2005power}
that the evidence leading to a suggestion of a power law in the
distribution is visual. Once we have these tests, the study goes
through the adjustment of a possible power law
\cite{merelo2017self,arafat2009commit}.

The general trend being the lack of statistical evidence that supports
the hypothesis of the existence of a power law, the truth is that as
described in \cite{newman2005power, clauset2009power} the most common
method for making subsequent adjustments, Leas Square Medium
\cite{merelo2017self,arafat2009commit,merelo16:self}, usually implies
the existence of a greater error than the possible alternatives.

Currently, the general trend that started with the description of the
power laws in self-organized systems has been slowed down. Instead,
due to the existence of more objective evaluation methods, the results
that were assumed until now are being re-evaluated.

Examples of this trend are papers such as \cite{Holme2019,
  Broido2019}, where the power laws in the networks and in the results
have been reevaluated, obtaining an analysis that can change the way
we make our assumptions about power laws.

Next we will present the methodology used to choose those repositories
and mine their information.


\section{Methodology}
\label{sec:method}

In this part we are going to explain the methods used in our
research. It is divided in four sections, explaining all procedures
used from the data extraction to the visualization. We emphasize the
second and third, being the ones that really represent a conceptual
change in the analysis of the existence of power laws in code
repositories.

\subsection{Data extraction}
We have chosen 16 repositories in different states of development.
These repositories were chosen in \cite{Merelo2016:repomining}, for
several reasons: They all represent a wide array of languages and
functionalities, from web frameworks such as Django to Atom plugins,
through one-of-a-kind frameworks such as Docker.

Normally a code repository is related to a software project, for this
reason, it is usual that they include several different languages
which are used in different parts of the project.  This mixture of
languages also offer a big range of variability between languages that
are interpreted or compiled, either to machine code or to bytecode.

Repositories vary also in {\em professionalism}, that is, the team
behind that software project. From a small Atom editor plugin to
TensorFlow, an open library created and maintained by a fully
professional community.

Repository mining was done during the months of January to March 2017,
with the second sweep of the same repositories performed in February
2019.

The way we look at changes in the repository was initially proposed by
\cite{Merelo2016:repomining} and was also used in
\cite{merelo2017self}, where a deeper explanation can be found.

This procedure is based on three main concepts:
\begin{itemize}
\item The usage of a discrete timeline formed by the commits, with
  every commit counting as time=1.
\item Work with selected files in the repository, excluding those
  related to images or style
\item We take the largest value from the inserted and deleted lines of
  code.
\end{itemize}

Please note that, in principle, we are not interested in the number of developers that participate in the project. A certain amount of developers is not a necessary condition for self-organization, but in any case we are interested in detecting power laws in these time series of change sizes in repositories, which do not include data at all on the number of users. The repositories we have been analyzing here do have a wide range of users, from one or two to several dozens but, as indicated, our methodology does not need to include information on the number of users.

Up to this point, there is nothing new. But is the way that we
evaluate our results that differs from previous analysis.

\subsection{Hypothesis tests}

Once the information from the repositories has been extracted, we
proceed to analyze it in order to find clues about what kind of
distribution is generating our data.

With that particular objective in mind, we change the methodology used
in \cite{merelo2017self} for several reasons.

First of all we want to offer a test that checks whether the observed
data set actually follows a power law, instead of only visualize the
result.  This kind of tests can vary based on different measured and
techniques. The one we are going to use is suggested in
\cite{clauset2009power}, which consists in using a goodness-of-fit
hypothesis test via bootstrapping procedure.

Due to the use of bootstrapping, this procedure is a time and
resources consuming one.  We essentially generate multiple data sets
with our two main parameters $x_{min}$ and $\alpha$ and then we
re-infer the model parameters. The outcome of the algorithm will
result in a P-value, which, if is large enough, tell us that the
difference between our data and the power law model we have generated
is small and mostly attributed to statistically randomness. On the
other hand, if P-value is close to 0, it is quite unlikely that our
model fits the data properly.  Following \cite{clauset2009power} we
choose our threshold value in $p=0.01$, in addition we are going to
perform this analysis with R package: {\tt PoweRlaw}. A detailed description
of it and the hypothesis tests can be found in
\cite{gillespie2015power}.

Up to this point, we have detected what data sets are unlikely to be
fitted by a power law in any of its range. Notice, on the contrary
case, that we are only assuming that we can not discard that a power
law is generating the data.  However, as there is some probability of
the appearance of this distribution in our data we can study and
analyze how its two main parameters may evolve between 2017 and 2019.

As it is unrealistic to think that a power law distribution will fit
all our data, our first step is to check what portion of the data
could be fitted with a power law, or in other words, what is the
minimal value (if there is one) from which the scaling relationship of
the power law begins \footnote{This is a fair assumption since we are
  working with heavy-tailed distributions and our main interest is the
  behaviour of the tail of our data.}.  This value is usually noted by
$x_{min}$ and is our first parameter.

Once we have the first value, we proceed to estimate the scale of our
power law.  As it is shown in papers like \cite{newman2005power,
  clauset2009power}, least square method is a poor but wide-spread
way to proceed when estimating the scale parameter. Instead, we are
going to use a direct method describe in \cite{clauset2009power} and
implemented in \cite{alstott2014powerlaw} that use the data values we
have. This method is known to produce a very nice fit with less error
than the others mentioned above.

Up to this point we have revisited our way of analyze power law
fitting and the estimation of our parameters. However, there is a more
deep question unanswered: does our data really follow (in a
statistically relevant way) a power law?  This kind of answers were
lacking in \cite{merelo2017self} and they are relevant independently
our first test's results, since they usually offer an unbiased look of
the data .

Taking into account that our main question is whether a power law is
the best description of our data, we choose to apply a comparative
test that could evaluate if there are any alternative distribution
that could have generated our data with greater likelihood than a
power law. That is the main reason why we choose to use a
log-likelihood ratio test implemented in \cite{alstott2014powerlaw}.
There are two algorithm's outcomes. First we have the log-likelihood
ratio between the two candidate distributions. The sign of this
quantity will point out which distribution is more likely to be
producing out data. After it, we calculate the signification of this
ratio, a P-value. Following \cite{alstott2014powerlaw} indications we
establish our P-value threshold at $p=0.05$; above that point the
loglikehood ratio has no significance and we can not decide which
distribution is better fitting out data.

\subsection{Classification}

With all this information we are able to offer a precise conclusion
about the probability that power law is generating our data. To sum up
all the tests in a single statement we use the scale proposed in
\cite{clauset2009power}, which is described as:
\begin{itemize}
\item \textbf{None}: Data-set is probably not distributed by a power
  law (first test failed).
\item \textbf{Moderate}: Power law is a possible fit but there are
  other plausible distributions that fit the data.
\item \textbf{Good}: Power law is a possible fit and none of the other
  distributions is plausible.
\item \textbf{Truncated}: when truncated power law is clearly favored
  over a simple power law
\end{itemize}

A relevant remark should be made: Even when the first test give us a
low probability of our data being distributed by a power law, the next
tests offers us an interesting insight. A power law can exceed others
distributions at explaining how our data is distributed, meaning that,
even though our data is not distributed by a power law, this fit could
offer us more information about our data than other distributions.

\subsection{Visualization}
 
On the visual aspects, a clarification should be made. Here we offer a
graphical view that differs from \cite{merelo2017self} but not a novel
one. It has been used in \cite{arafat2009commit} with similar
purposes.  Briefly, we are going to use the probability density
function (PDF) for plotting. Due to the requirement of binning the
data to this type of graphic, we are going to use a logarithmic
spacing, since it reduces the statistical errors in the tail in
log-log plots at it is stated in \cite{newman2005power}.


\section{Results}
\label{res}
\begin{table*}[h!tbp]
  \caption{2017 Parameters with power law fitting}
  \begin{center}
    \begin{tabular}{| p{0.12\linewidth} | p{0.1\linewidth} |
        p{0.3\linewidth} | p{0.3\linewidth} |}
      \hline
      Repository name &$x_{min}$ & $\alpha$ & $\sigma$\\
      \hline
      atom &10.0 &2.039242998376567  &0.15492119930071382 \\
      cask &18.0 &1.7336857967225443  &0.06725686671465021 \\
      Dancer2 &107.0 &2.177500075690754  &0.07730666731550309 \\
      django &10786.0 &2.1744317304939558  &0.01917838984962695 \\
      docker &73.0 &1.4645926562553306  &0.004754872537365103\\
      ejabberd &2.0 &1.228631569923214  &0.0033017325614456455 \\
      fission &25.0 &1.7697310637483803  &0.09937185303186655 \\
      mojo &32.0 &2.1283668081735456  &0.03162540535355575 \\
      Moose &36.0 &1.768059278495766  &0.0197132783524576 \\
      rakudo &11.0 &1.6159534004325067  &0.006834230629556248 \\
      scalatra &105.0 &1.7126278313239238  &0.02583273870367301 \\
      tensorflow &6.0 &1.2136744934008  &0.002587381661200525 \\
      tpot &39.0 &1.601798428793051 &0.0374663336679551 \\
      tty &85.0 &2.5687968664786887  &0.16355837969418938 \\
      vue &10.0 &1.4008340778667951  &0.007381204719143647\\
      webpack &66.0 &1.7525127838915868 &0.030568742256899546\\ 
      \hline
    \end{tabular}
  \end{center}
  \label{tab:2017pars}
\end{table*}
\begin{table*}[h!tbp]
  \caption{2019 Parameters with power law fitting}
  \begin{center}
    \begin{tabular}{| p{0.12\linewidth} | p{0.1\linewidth} |
        p{0.3\linewidth} | p{0.3\linewidth} |}
      \hline
      Repository name &$x_{min}$ & $\alpha$ & $\sigma$\\
      \hline
      atom  &9.0  &1.8359195263285408  &0.11705214645679832 \\
      cask  &22.0   &1.7974451838770937  &0.07249501671609943 \\
      Dancer2  &106.0   &2.261037545006354  &0.08425661509414822 \\
      django  &22380.0   &2.6887068697480156  &0.03580856180986959\\
      docker  &85.0  &1.471139237773135  &0.004450258248742491 \\
      ejabberd  &4.0   &1.2671552485187476  &0.00372055095134032 \\
      fission  &22.0   &1.4589753590316525  &0.021399836458042016\\ 
      mojo  &32.0   &2.1384964656994434  &0.031043490328162827 \\
      Moose  &36.0   &1.7677961781806713  &0.019622681314368864 \\
      rakudo  &12.0   &1.603181925304308  &0.006028504486885445 \\
      scalatra  &105.0  &1.678355887867916  &0.02363164712583398 \\
      tensorflow  &16.0  & 1.276851335699674  & 0.0017743964236792348 \\
      tpot  &176.0   &2.139617006349855  &0.08401368034180524 \\
      tty  &83.0   &2.6669971731409587  &0.1582243695783554 \\
      vue  &12.0   &1.4407170088980175  &0.007474075257736019 \\
      webpack &50.0  &1.481958114820392  &0.00906295021690492 \\
      \hline
    \end{tabular}
  \end{center}
  \label{tab:2019pars}
\end{table*}
First, analyzing the parameters of the possible power law
distribution, we find that in those repositories with low variation of
the $x_{min}$ parameter there is a lack of correlation between the
changes in the estimated starting point from which the distribution of
power law starts and the scale parameter, $\alpha$. Even if small
changes in $xmin$ produce small changes in $\alpha$, the lack of
correlation probably means that this changes are the result of the
natural stochasticity of the evolution process.

However, as it can be seen in Tables \ref{tab:2017pars} and
\ref{tab:2019pars}, repositories such as Django, which have a strong
variation in the initial point, also have a greater variation in the
scale parameter. This is not produced because of a global change in the overall
distribution. The main reason for this change is the variation of the $x_{min}$ parameter, due to the existence of a different range of commits that fits a power law better than the previous one. 
\begin{table*}[h!tbp]
  \caption{2017 summary results}
  \begin{center}
    \begin{tabular}{| c |c| c | c| c |c | c |}
      \hline
      Repository name & KS test & PL vs. LogN & PL vs. Exp & PL vs. PLtrunc & Result \\ 
      \hline
      atom & 0.244 &ND & ND &PLtrunc & Moderated \\ 
      cask & 0.236 &ND & PL & ND & Moderated \\
      Dancer2 &0.178 &ND &PL &ND & Moderated \\
      django &0.002 &LogN &PL &PLtrunc &None\\
      docker &0 &LogN &PL &PLtrunc &None\\
      ejabberd &0 &LogN &PL &PLtrunc &None\\
      fission &0.654 &ND &PL &PLtrunc & Moderated\\
      mojo &0.874 &PL &PL &ND & Moderated \\
      Moose &0.2 &PL &PL &PLtrunc & Truncated\\
      rakudo &0 &PL &PL &PLtrunc & None\\
      scalatra &0 &LogN &PL &PLtrunc & None\\
      tensorflow &0 &LogN &PL &PLtrunc & None\\
      tpot &0.01 &LogN &PL &PLtrunc &None\\
      tty &0.418 &ND &ND &ND & Moderated \\
      vue &0 &PL &PL &PLtrunc & None\\
      webpack &0.004 &LogN &PL & PLtrunc & None\\
      \hline
    \end{tabular}
  \end{center}
  \label{tab:2017tests}
\end{table*}
\begin{table*}[h!tbp]
  \caption{2019 summary results}
  \begin{center}
    \begin{tabular}{| p{0.12\linewidth} | p{0.08\linewidth} |
        p{0.08\linewidth} | p{0.08\linewidth} | p{0.1\linewidth}
        |p{0.13\linewidth} | p{0.09\linewidth} |}
      \hline
      Repository name & KS test & PL vs. LogN & PL vs. Exp & PL vs. PLtrunc & Result \\ 
      \hline
      atom &0.108 &ND &ND &PLtrunc & Moderate\\
      cask &0.276 &ND &PL &ND & Moderate \\
      Dancer2 &0.426 &ND &PL &ND & Moderate \\
      django &0.862 &LogN &PL &PLtrunc &Truncated \\
      docker &0.118 &LogN &PL &PLtrunc & Moderate \\
      ejabberd &0 &LogN &PL &PLtrunc & None\\
      fission &0 &ND &PL &PLtrunc & None \\
      mojo &0.88 &PL &PL &ND &Moderate \\
      Moose &0.224 &PL &PL &PLtrunc &Truncated \\
      rakudo &0 &PL &PL &PLtrunc & None \\
      scalatra &0 &LogN &PL &PLtrunc & None \\
      tensorflow &0.418 &LogN &PL &PLtrunc & Moderate \\
      tpot &0.184 &LogN &PL &PLtrunc & Moderate \\
      tty &0.47 &ND &ND &ND & Moderate \\
      vue &0 &PL &PL &PLtrunc & None \\
      webpack &0.002 &LogN &PL &PLtrunc & None\\
      \hline
    \end{tabular}
  \end{center}
  \label{tab:2019tests}
\end{table*}
Another special remark should be addressed to Tensorflow. The
distribution fit of the repository in 2019 has caused a little
controversy between \cite{alstott2014powerlaw} and
\cite{gillespie2015power}, as the final numerical results where different.
 We believe the main reason for that is the
two different sections detected by their algorithms. One package gives
more probability of fitting a power law between 10 to $10^5$ commit
size and the other starting from $10^5$ commit size till the
end. However, choosing different starting points produced similar
results in the final tests. 

After analyzing the parameters estimated by assuming a power law fit,
it is time to check what are 
the results from the hypothesis tests, made to know if there are other
possible distributions generating the data. These results can be found
in Tables \ref{tab:2017tests} and \ref{tab:2019tests}.

We can observe that in general, as the repositories evolve over time,
there is a tendency to the appearance of a greater probability in
favor of power law existence, as None cases have been reduced between
2017 and 2019.

Even so, it is worth noting the complete lack of repositories in which
our confidence in finding a power law distribution is high (following
our scale: \textit{Good}).

\begin{figure*}[htbp]
  \centerline{\includegraphics[width=0.95\textwidth]{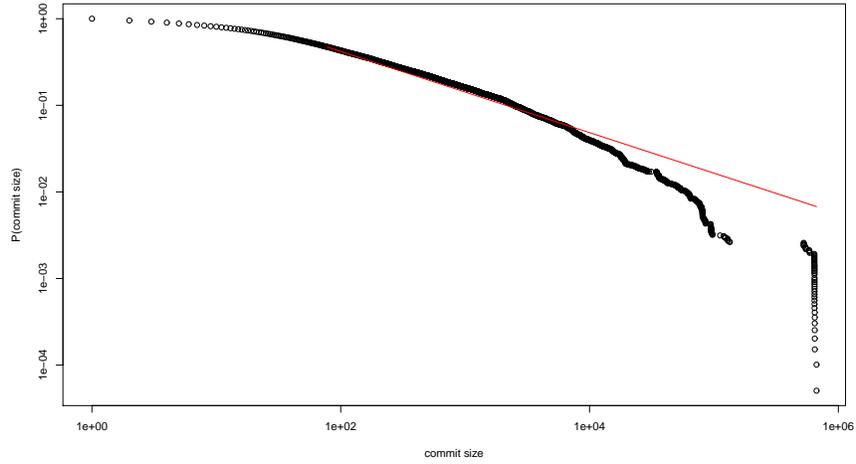}}
  \caption{PDF plot of a repository with \textit{None} rank: Docker
    from 2017.}
  \label{fig:docker}
\end{figure*}

Faced with the most common visualizations, the one used in our paper
highlights even more how we can not affirm the existence of power laws
distributions if we look at them in detail. See as an example how
Fig. \ref{fig:docker}. shows that the adjustment of a power law would
not capture some characteristics in the tail of our data.

\begin{figure*}[htbp]
  \centerline{\includegraphics[width=1\textwidth]{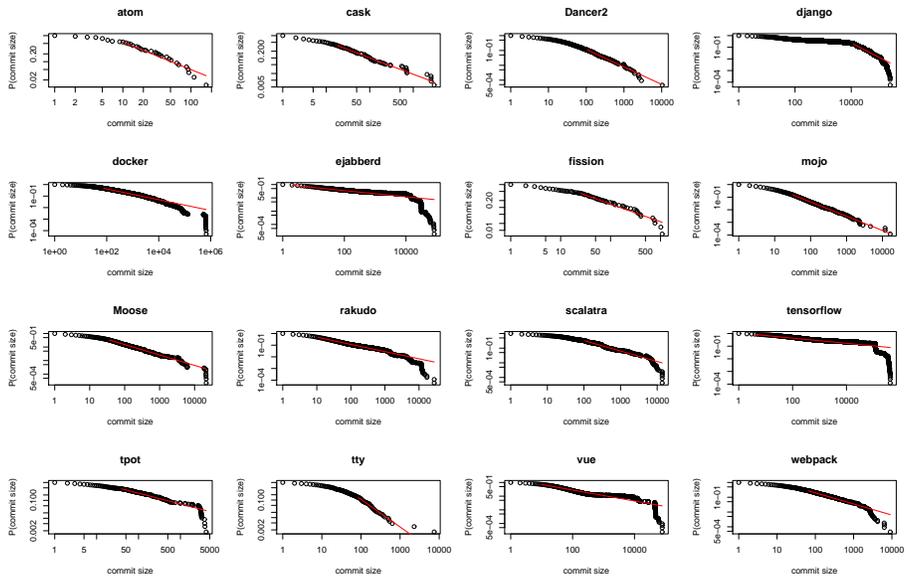}}
  \caption{PDF plot of all repositories 2017. Red line represents the
    power law fit and the area where it fits.}
  \label{fig:pl2017}
\end{figure*}

\begin{figure*}[htbp]
  \centerline{\includegraphics[width=1\textwidth]{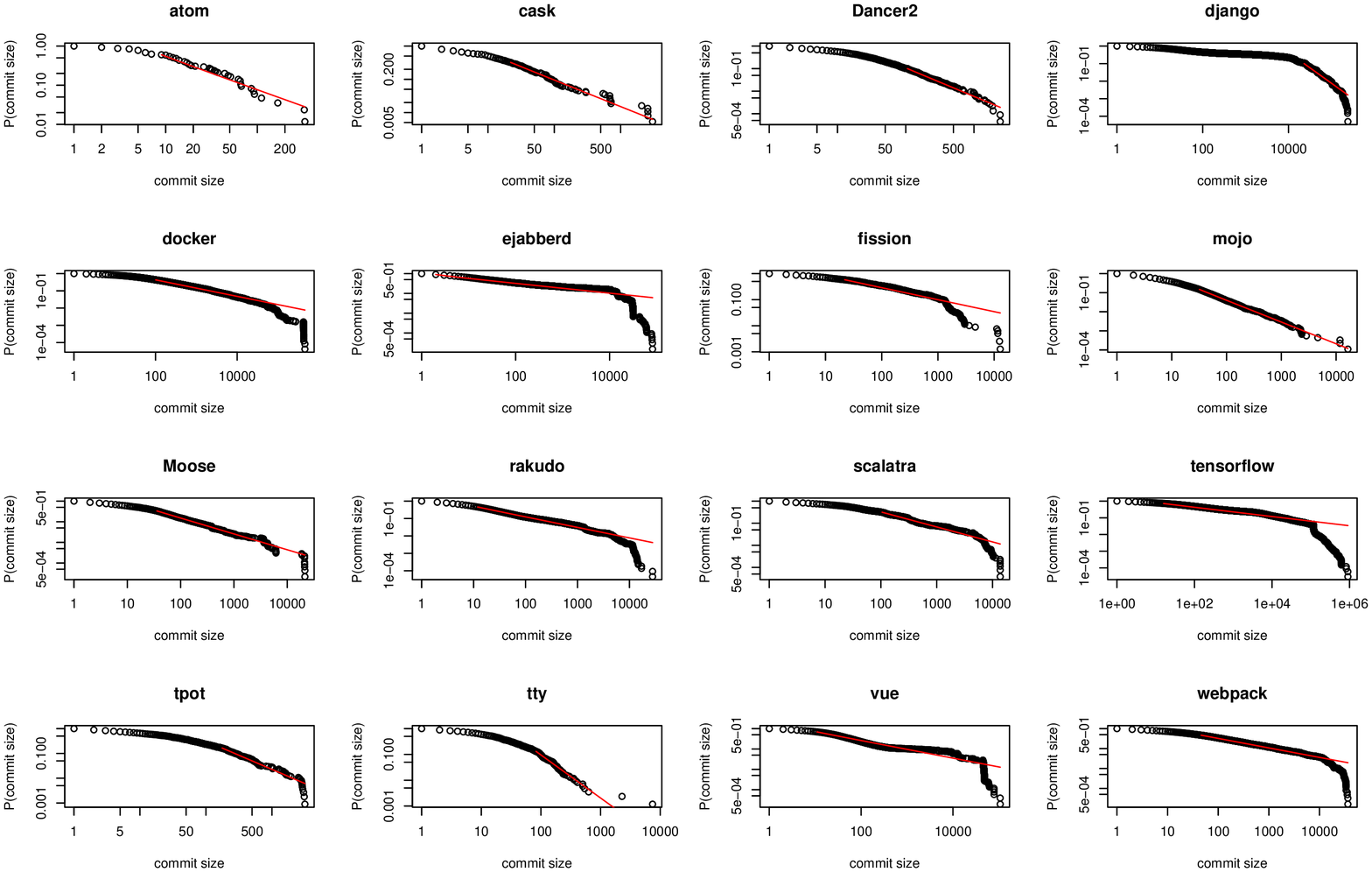}}
  \caption{PDF plot of all repositories 2019. Red line represents the
    power law fit and the area where it actually fits.}
  \label{fig:pl2019}
\end{figure*}

We would like to remark that, even in smaller plots like
\ref{fig:pl2017} and \ref{fig:pl2019}, where we could easily say that
there are power laws only by sight, the mathematics presented in this
paper support that the probability of that happening is, as maximum,
\textit{ moderate}.

\begin{figure*}[htbp]
  \centerline{\includegraphics[width=0.95\textwidth]{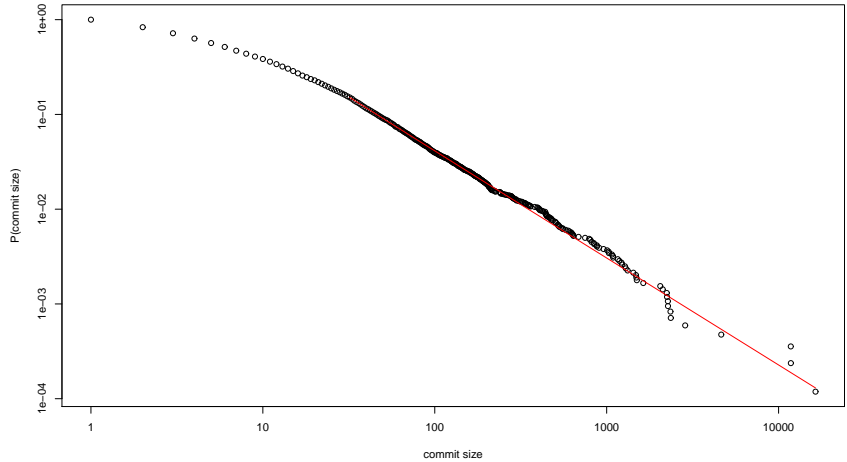}}
  \caption{PDF plot of a repository with \textit{Moderate} rank: Mojo
    from 2017.}
  \label{fig:mojo}
\end{figure*}

A proper example of this kind of biased assumptions is
Fig. \ref{fig:mojo}. Seen the Mojo distribution and power law fit we
can suspect the existence of a power law in the repository. Its
distribution that also remains in time till 2019. This can be an
indicative that this repository is self-organizing to stay in that
point. But even in this case, the probability of a power law is only,
again, \textit{moderate}.

On overall the higher rank is not achieved in any repository. We can only 
state that in most of the repositories a power law could explain the data, 
but, other distributions are equally probable.

\section{Conclusions}\label{conc}

In general, we should always use a standard mathematical approach in
order to affirm any behaviour in the distribution of our data. That way, we
can be assured that our results are unbiased, not simply based in our sight or
plots, and they have been tested statistically. This is what we have
done in this paper, that not only uses that methodology on data already published,
but also updates the data checking whether there has been any
evolution in the actual existence of a power law in the repositories
under study.

In general, the probability of finding a power law in the repositories
studied is quite low, so, up to now, we can not categorically affirm 
that the studied systems are in a critical state due to
self-organization, at least from this point of view that uses the size of changes in every commit in a repository.

This result, however, does not imply that there is not any kind of
self-organization or even a power law, but that the systems have not
been regulated enough and, therefore has not been established near a
possible critical state.

Furthermore we would like to explore the possible causes of such
dramatic changes as the ones seen in repositories like Django, since
they may be related to a possible phase change. 

In any case, the article presents an principled and objective workflow
to study the existence of power law distributions that settles the
status of these kind of measures in software development via
repositories.

Thanks to this procedure, we hope we can establish a new state of
art, from which researchers can begin to study these and other
properties of code repositories in order to understand which of them
may be related to self-organization and which ones have a very low
probability of that.

This paper is also a starting point for the in-depth study of how,
when and what produces complex behaviors or self-organized criticality
in code repositories. We have only treated the number of commits, but
there are many other parameters or variables (contributions per
author, per file, blame chunks sizes \footnote{Code lines that are next to
  each other and were modified together} or authors in different
files) that can cause the appearance of power laws within the
repositories, which could give us clues about a possible critical
state of our system.

On the other hand, as our sample size of repositories is relatively small, we could
extend this study to different repositories, to detect possible
general patterns or patterns related to repositories written in
certain languages or ascribed to certain software projects.

We would like to conclude pointing out that all the software used in
this research (being the main ones R\cite{R}, Python \cite{CS-R9526}
and \LaTeX) is open source. Furthermore, all the scripts used and data
obtained (curated and not curated) are open and can be checked at
\url{https://github.com/geneura-papers/2019-SASO-Repos-Powerlaws}.

\bibliographystyle{apalike}
\bibliography{analyzing-repos-saso19}

\end{document}